\documentclass[aps,prl,twocolumn,superscriptaddress]{revtex4}

\usepackage{graphicx}
\usepackage{amsmath}
\graphicspath{{figures/}}
\usepackage{xcolor}
\usepackage[normalem]{ulem}

\begin{document}


\title{High order magnon bound states in the quasi-one-dimensional antiferromagnet ${\alpha}$-NaMnO$_2$}


\author{Rebecca L. Dally}
\altaffiliation{Contributed equally to this work}
\affiliation{NIST Center for Neutron Research, National Institute of Standards and Technology, Gaithersburg, MD 20899, USA}
\affiliation{Materials Department, University of California, Santa Barbara, CA 93106, USA}

\author{Alvin J.R. Heng}
\altaffiliation{Contributed equally to this work}
\affiliation{Kavli Institute for Theoretical Physics, University of
	California, Santa Barbara, Santa Barbara, CA 93106, USA}
\affiliation{Division of Physics and Applied Physics, School of Physical and Mathematical Sciences, Nanyang Technological University, Singapore 637371, Singapore}

\author{Anna Keselman}
\affiliation{Kavli Institute for Theoretical Physics, University of
	California, Santa Barbara, Santa Barbara, CA 93106, USA}

\author{Mitchell M. Bordelon}
\affiliation{Materials Department, University of California, Santa Barbara, CA 93106, USA}

\author{Matthew B. Stone}
\affiliation{Neutron Scattering Division, Oak Ridge National Laboratory, Oak Ridge, TN 37831, USA}

\author{Leon Balents}
\email[]{balents@kitp.uscb.edu}
\affiliation{Kavli Institute for Theoretical Physics, University of California, Santa Barbara, Santa Barbara, CA 93106, USA}

\author{Stephen D. Wilson}
\email[]{stephendwilson@uscb.edu}
\affiliation{Materials Department, University of California, Santa Barbara, CA 93106, USA}

\date{\today}

\begin{abstract}
Here we report on the formation of two and three magnon bound states in the quasi-one-dimensional antiferromagnet $\alpha$-NaMnO$_2$, where the single-ion, uniaxial anisotropy inherent to the Mn$^{3+}$ ions in this material provides a binding mechanism capable of stabilizing higher order magnon bound states.  While such states have long remained elusive in studies of antiferromagnetic chains, neutron scattering data presented here demonstrate that higher order $n>2$ composite magnons exist, and, specifically, that a weak three-magnon bound state is detected below the antiferromagnetic ordering transition of NaMnO$_2$.  We corroborate our findings with exact numerical simulations of a one-dimensional Heisenberg chain with easy-axis anisotropy using matrix-product state techniques, finding a good quantitative agreement with the experiment.  These results establish $\alpha$-NaMnO$_2$ as a unique platform for exploring the dynamics of composite magnon states inherent to a classical antiferromagnetic spin chain with Ising-like single ion anisotropy.    
\end{abstract}

\maketitle

\section{\label{intro}Introduction}

One dimensional systems are renown for their ability to host ground states and phases markedly different from their higher dimensional counterparts.  In the realm of magnetism, the prototypical $S=1/2$ Heisenberg antiferromagnetic chain realizes a phase without order and power-law correlations, described at low energies by conformal field theory \cite{haldane1993exact, PhysRevLett.73.332}. The corresponding $S=1$ chain also avoids order, and hosts instead a topologically non-trivial Haldane-gapped state with protected $S=1/2$ boundary spins \cite{PhysRevLett.50.1153, affleck1989quantum}.  While such quantum paramagnetic (i.e. not ordered) states in principle persist for any $S$ in one dimension in the ideal case, they become increasingly fragile to perturbations as $S$ increases and the semi-classical limit is achieved.  In this regime, described by the non-linear sigma model (NLSM) with a weak coupling of order $1/S$ \cite{haldane1983continuum}, quantum fluctuations are confined to very low energies. A small anisotropy restricts those fluctuations to the ``easy'' directions, and for an Ising-like situation, this is sufficient to induce order.

Thus, large spin (e.g. $S\geq 2$) quantum spin chains seem an unlikely place to observe strongly quantum phenomena.  Indeed this is true for their ground states, however, their excitations can still be highly quantum and realize interesting and paradigmatic few-body problems.  Recent interest in few-body problems -- i.e. the quantum mechanics of a finite number $n>2$ of interacting particles \cite{RevModPhys.58.361} -- stems largely from ultra-cold atoms, where examples include the formation of ``droplets'' of attractive bosons \cite{PhysRevA.97.011602}, and the Efimov states (three or more body bound states) in the unitary limit near a Feshbach resonance \cite{zaccanti2009observation}.  Generally this bound state formation is strongest in one dimension, and the multiple particle case is also known to be possible in spin chains from the study of field-induced multipolar orders in $J_1$-$J_2$ models \cite{PhysRevB.80.140402}.  In the latter case, the particles binding are magnons above a trivial field-induced polarized state \cite{PhysRevB.76.060407}, and lattice-scale competing interactions play an important role. 

The cleanest, arguably most beautiful example of a few-body problem is the droplet mentioned above, which is a collection of one-dimensional Bosons with attractive zero-range delta-function interactions.   The fundamental physics of this problem is that increasing numbers of bosons bind more strongly, due to Bose statistics.  Moreover, the ground state is, in this case, exactly soluble analytically via a simple application of the Bethe ansatz.   In prior work, we uncovered a surprising connection of this canonical model problem to the large $S$ nearest-neighbor antiferromagnetic chain with weak easy-axis Ising anisotropy \cite{Dally_2018}.  The theory developed there predicts an approximate mapping of the quantum mechanics of magnons to the droplet problem of bosons. This mapping is non-trivial because the unperturbed NLSM is a gapless theory and, in it, binding would mean instability.  Binding occurs as a subtle balance between fluctuations, weak anisotropy, and the intrinsic interactions of the NLSM.  

In Ref. \citenum{Dally_2018}, the two-body bound state predicted by this theory was observed in the material $\alpha$-NaMnO$_2$. Fundamentally, the crystal lattice of $\alpha$-NaMnO$_2$ is built from two-dimensional sheets of triangular-lattice planes of Mn$^{3+}$ moments, yet the Jahn-Teller effect inherent to the Mn-cations drives a coherent distortion of the triangular lattice into coupled isosceles triangles comprised of two long Mn-Mn distances and one short Mn-Mn exchange pathway \cite{Giot_PRL_2007}.  This breaks the frustration of the triangular lattice and defines a strong AF nearest neighbor exchange energy ($J_1$) along the short bond length that defines the AF chain direction.  Coupling between these chains is frustrated by equivalent AF next nearest neighbor couplings ($J_2$) via the two longer legs of the triangular lattice.  The result is a highly one-dimensional spin system \cite{PhysRevB.77.024412,stock2009one} with effective $J_1/J_2=8$ \cite{Dally_2018} that also possesses a weak Ising-like single-ion anisotropy, $D$, which we will estimate to be $D/J_1 = 0.086$.  

\begin{figure}[t]  
	\centering
	\includegraphics[scale=0.325]{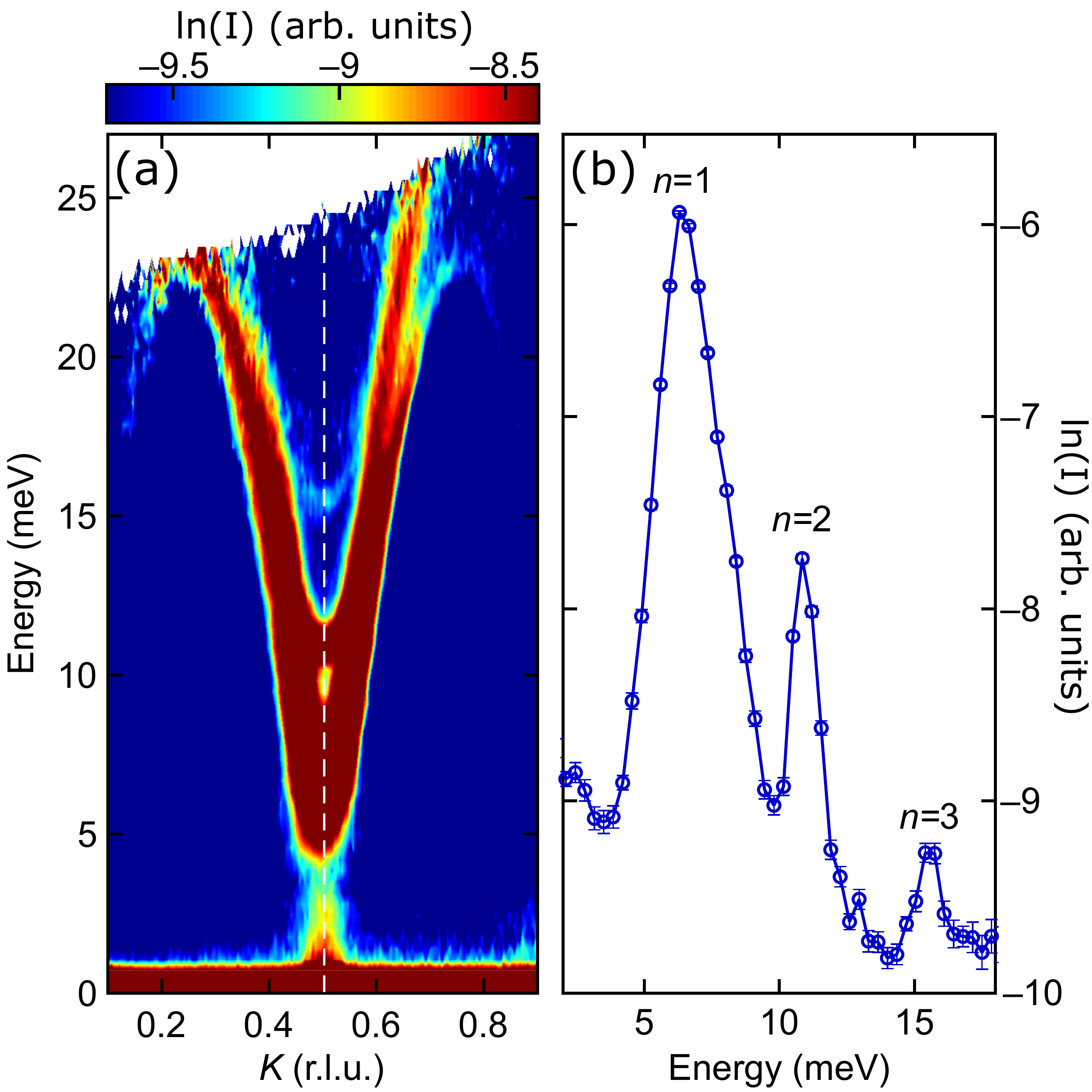}   
	\caption{Neutron scattering data collected with $E_i=30$ meV at $T=4$ K. The data were integrated throughout the entire zone in both $H$ and $L$. Panel (a) shows an intensity of the data on a logarithmic scale. The dashed, white line shows the direction of the energy cut plotted in panel (b). Panel (b) shows an energy cut through the intensity map centered at the AF zone center, $K=0.5$, with the $n=1$, $n=2$, and $n=3$ magnon modes.}
	\label{fig:twomagdisp}
\end{figure}

\begin{figure}[t]  
	\centering
	\includegraphics[scale=0.175]{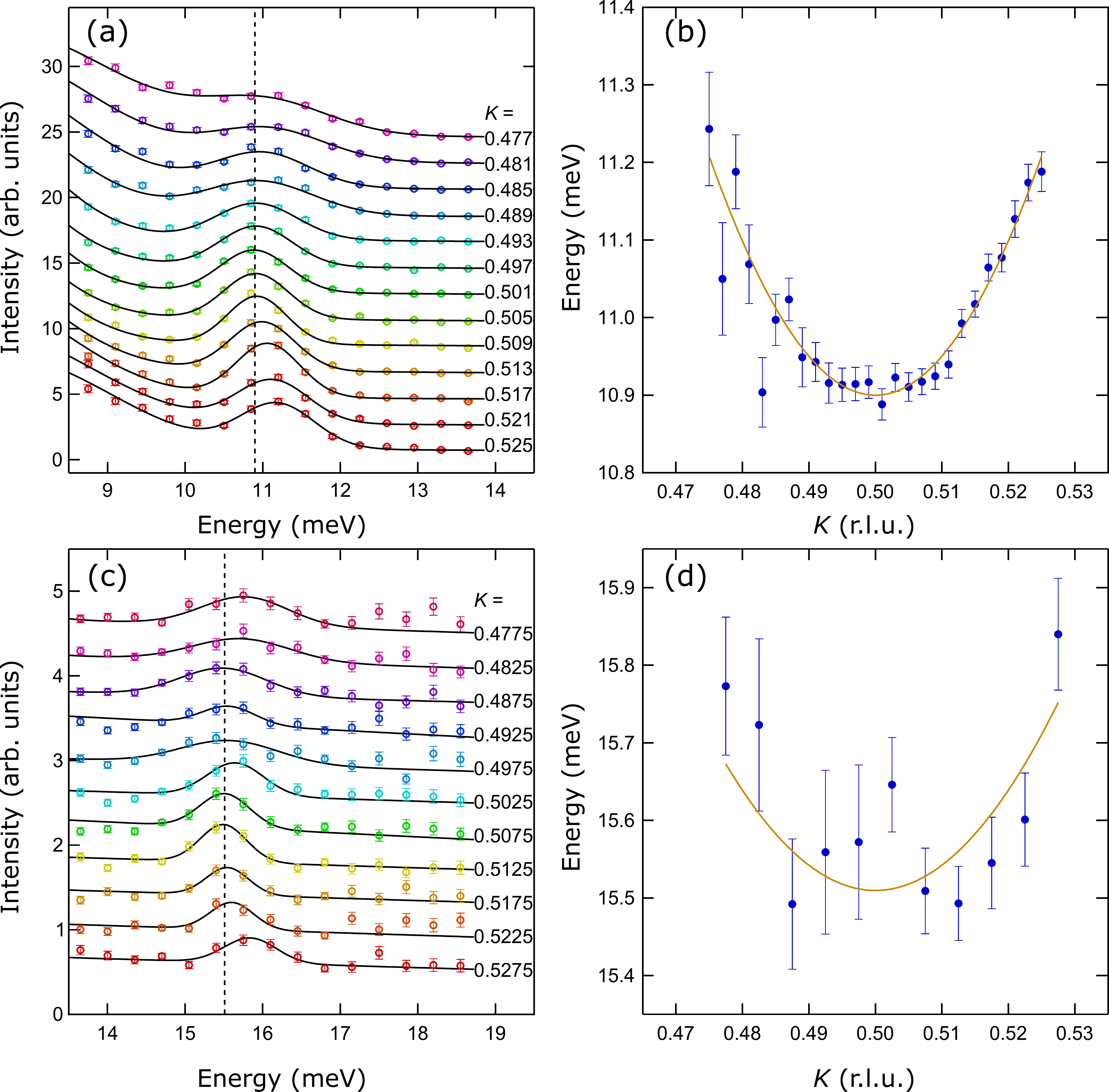}   
	\caption{Constant $K$ cuts parameterizing the dispersions of the $n=2$ and $n=3$ modes. The data were integrated throughout the entire zone in both $H$ and $L$ and represent cuts through the color plot in Fig. 1. (a) Individual $K$-cuts parameterizing the dispersion of the $n=2$ mode. (b) The fit energies of the $n=2$ mode plotted as a function of $K$. (c) Individual $K$-cuts parameterizing the dispersion of the $n=3$ mode (d) The energies of the $n=3$ mode as a function of $K$. The K cuts in panels (a) and (c) are offset from one another for clarity, and the solid black lines are fits parameterizing the modes' dispersions as described in the text. The solid orange lines in panels (b) and (d) are fits to the 1D dispersion described in the text.}
	\label{fig:fifteendisp}
\end{figure}

In this paper, we present an acid test of the mapping of the spin lattice of $\alpha$-NaMnO$_2$ to the droplet problem by uncovering a three-body bound state in $\alpha$-NaMnO$_2$. Both $n=2$ and $n=3$ magnon bound states are observed in inelastic neutron scattering measurements with binding energies consistent with a minimal model of AF spin chains possessing an easy-axis uniaxial, single-ion anisotropy.   Notably, the proper theory requires consideration of the ``charge" of the magnons that bind, which breaks integrability and leads to a reduction of the binding energy.   Our modeling suggests $\alpha$-NaMnO$_2$ and other weakly anisotropic, large $S$ chains can be excellent test platforms for few-body physics.

A single crystal of ${\alpha}$-NaMnO$_2$ was grown via the floating zone method \cite{DALLY2017203}, and neutron scattering measurements were performed on the time-of-flight neutron spectrometer SEQUOIA \cite{Granroth_2010} at the Spallation Neutron Source at Oak Ridge National Laboratory. The crystal was sealed in a cryostat with 4He exchange gas, and the $(0, 1, 0)$ and $(1, 0, 1)$ crystallographic directions were aligned in the horizontal scattering plane. Data were collected with incident neutron energies of $E_i = 30$ and $60$ meV with the Fermi chopper in high-resolution mode, and the sample was rotated about the $(-1, 0, 1)$ axis in $1 ^{\circ}$ increments over a range of $180 ^{\circ}$. Data were reduced and analyzed using the software package Horace \cite{Horace}. Throughout the paper, positions in momentum space, $\mathbf{Q}$, are reported in reciprocal lattice units (r.l.u.), where $H$, $K$, and $L$ reflect $\mathbf{Q}[{\AA}^{-1}] = (\frac{2 \pi}{a\sin{\beta}}H, \frac{2\pi}{b}K, \frac{2\pi}{c\sin{\beta}}L)$ with $a=5.63$ \AA, $b=2.86$ \AA, $c=5.77$ \AA, and $\alpha=\gamma=90^{\circ}; \beta=113^{\circ}$. 

Fig.~\ref{fig:twomagdisp} shows inelastic neutron scattering data collected about the 1D AF zone center, (0, 0.5, 0), in the AF ordered state ($T=4$ K). As the spin dynamics in this system are quasi-1D, data are integrated across the entire zone in $H$ (the interchain) and $L$ (the interplane) directions. The resulting color map of intensities plotted in Fig. 1 (a) shows three branches of excitations dispersing along $K$ (the intrachain direction), each of which are centered at the AF zone center $K=0.5$ position. Fig. 1 (b) shows an energy cut through this map centered at the AF zone center.  Looking along this cut, the lowest energy mode appears at the expected zone center magnon gap near $E_1=6.15$ meV and is comprised of 4 nearly degenerate modes superimposed from the two crystallographic and two magnetic twins inherent to this material.  A second mode, which is purely 1D, is seen centered at $E_2=10.9$ meV and matches the energy of the previously reported, longitudinally polarized spin excitation identified as a two-magnon ($n=2$) bound state. Remarkably, an additional 1D mode appears at higher energy at $E_3=$15.5 meV---this third mode is newly resolved and suggests the formation of a higher order three-magnon ($n=3$) bound state.

The dispersions of the $n=2$ and $n=3$ bound states are illustrated about the AF zone center via constant $K$ cuts (offset from one another for clarity) and plotted in Figs. 2 (a) and (c).  Fits parameterizing the energies of each mode as a function of $K$ were performed using Gaussian peaks on a sloping background with the resulting dispersion relations shown in Figs. 2 (b) and (d).   These relations were then empirically quantified via a fit to the form $E(K)=\sqrt{{\Delta^2} + c^2\sin^2{2{\pi}K}}$ with $\Delta$ reflective of the gap energy and $c$ an empirical spin stiffness parameter.

For the $n=2$ branch of excitations, the results of the fit are shown in Fig. 2 (b), where ${\Delta}$ and $c$ were found to be $10.900 \pm 0.006$ meV and $16.7 \pm 0.4$ meV, respectively.  The dispersion of the higher energy $n=3$ branch shown in Fig. 2 (d) yields ${\Delta} = 15.51 \pm 0.03$ meV and $c = 16 \pm 2$ meV. We note here, that while clear dispersion is evident in the data, parameterizing the effective spin-stiffness parameter for the $n=3$ branch is coupled to the $K$ bin size chosen. This is due to the weak nature of $n=3$ mode (which is nearly two orders of magnitude weaker than the $n=1$ peak). Using the parameterization for $n=3$ mode shown in Fig. 2 (d), the effective masses of the modes $m_i\propto\Delta/c^2$ have ratios of $m_1:m_2:m_3$=$1:3.5:5.4$.

The temperature evolution of these modes are shown in Fig. 3. Energy cuts along the 1D AF zone center are plotted at three different temperatures: $T = 4$ K (in the AF ordered state), $30$ K (in the incommensurate short-range ordered state), and $50$ K (in the high temperature regime of quasi-one dimensional correlations) \cite{Dally_2018_PRB}. True long-range order along the chains (divergent $K$-axis correlation lengths) occurs only below $T=22$ K \cite{Dally_2018_PRB}, and both $n=2$ and $n=3$ modes vanish above this temperature. The $n=1$ single magnon peak persists to high temperature as it becomes increasingly damped and broadens into the single-ion anisotropy gap with increasing temperature.

\begin{figure}[t]  
\centering
\includegraphics[scale=0.45]{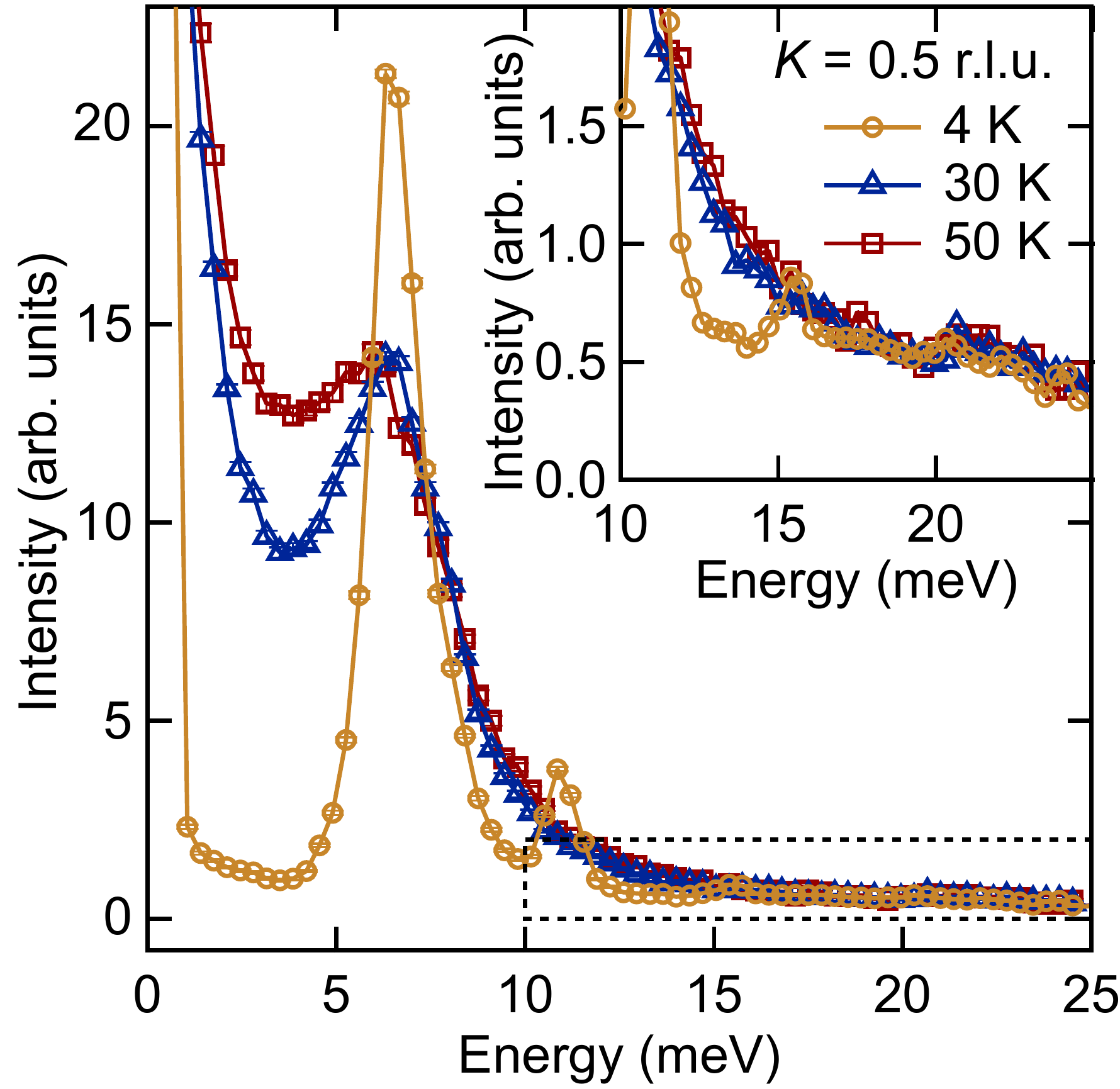}   
\caption{$E_i=30$ meV data showing the temperature dependence of spin excitations determined via an energy cut at the quasi-1D zone center, $\mathbf{Q}=(0,0.5,0)$.  The inset shows a zoomed in region of the data, highlighting the temperature dependence of the $n=3$ mode. Data are integrated across a width of 0.01 r.l.u. in $K$.}
\label{fig:sixallT}
\end{figure}

We now proceed to interpret the above results theoretically.  Prior work on this compound developed a semi-classical theory to leading order in $1/S$ in the quasi-1D limit \cite{Dally_2018}.  In that theory, $1/S$ corrections induce an attractive delta-function-like interaction between magnons, which creates the $n=2$ two magnon bound state.  Within the same theory higher bound states are expected, as described in the supplementary materials \cite{supplemental}.  Here to obtain a more quantitative comparison with the experimental data that does not rely on the $1/S$ expansion, we carried out numerically exact matrix product state (MPS)-based~\cite{Schollwoeck2011} calculations in the one dimensional limit using the ITensor library~\cite{ITensor}.

Given the quasi-1D nature of $\alpha$-NaMnO$_2$, in our numerical simulation, we consider a $S=2$ antiferromagnetic Heisenberg chain with single-ion anisotropy
\begin{equation}
    \label{eq:hamiltonian}
    H = J_1 \sum_j \vec{S}_j \cdot \vec{S}_{j+1} - D \sum_j (S_j^z)^2
\end{equation}
where the sum over $j$ indicates a sum over spins on a 1D lattice. We calculate the spectral function, which at zero temperature is given by
\begin{equation}
\label{eq:spectral_func}
S(k, \omega)=\int_{-\infty}^{\infty} d t e^{i \omega t} \sum_{j=-\infty}^{\infty} e^{-i k j}\left\langle \vec{S}_{j}(t) \cdot \vec{S}_{0}(0)\right\rangle,
\end{equation}
where the expectation value is taken in the ground state of the system.  We consider a finite chain with $N=500$ sites and start by obtaining the ground state of the system using density matrix renormalization group (DMRG)~\cite{White1992}. We then perform time evolution up to times $t_{\rm max}=20 J_1^{-1}$ using time evolving block decimation (TEBD)~\cite{Vidal2003,Vidal2004,supplemental}.

Using this model, we obtain an independent estimate for the couplings $J_1$ and $D$ from a least squares fit of the dispersion of the $n=1$ mode, obtained from the time-of-flight neutron experiment. The single-magnon mode momentum slice along the $(H,H,0)$ direction (corresponding to a high-symmetry direction in the 3D reciprocal space) was modeled via the spectral function calculated numerically. This fit yields $J_1=5.34\pm0.08$meV and $D=0.46\pm0.02$meV, roughly consistent with earlier estimates obtained from a fit to linear spin wave theory (see supplementary materials for further discussion \cite{supplemental}).

\begin{figure}[t]  
\centering
\includegraphics[scale=0.7]{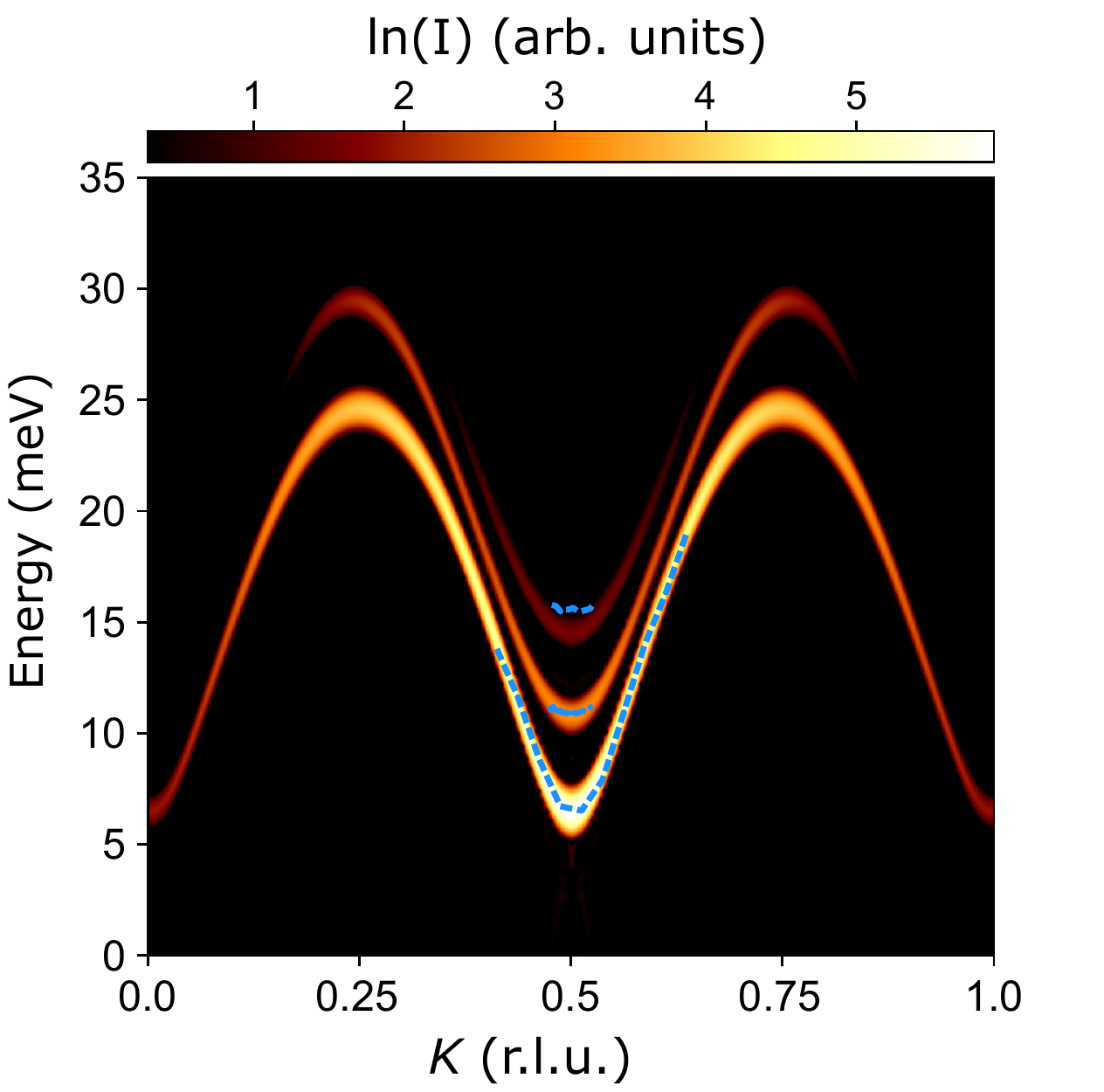}   
\caption{
Spectral function obtained numerically using the 1D model, for $J_1=5.34$ meV and $D=0.46$ meV. The intensity is normalized to a log scale.
Three modes are clearly seen in the vicinity of $K=0.5$.  Blue dashed lines plotted on top correspond to dispersions obtained from experiment. The $n=1$ data is obtained from a $(H, H, 0)$ cut, with integration in the $L$ direction from $-0.5<L \  (r.l.u) <0.5$ and integration in the $(-H, H, 0)$ direction from $-0.01<H \ (r.l.u)<0.01$. The $n=2$ and $n=3$ data were obtained via cuts about the AF zone center from Fig. \ref{fig:twomagdisp}(b) and Fig. \ref{fig:fifteendisp}(c) respectively. 
}
\label{fig:dmrg}
\end{figure}

The spectral function obtained numerically for these values of $J_1$
and $D$ is shown in Fig. \ref{fig:dmrg}, with the experimentally
measured dispersions for the three modes plotted on top of it as blue
dashed lines.  The $n=1$, $n=2$, and $n=3$ magnon modes are clearly
visible in the spectral function in the vicinity of $K=0.5$. The
corresponding gaps are given by $E_1=6.4\pm0.4$ meV, $E_2=10.7\pm0.4$
meV and $E_3=14.5\pm0.5$ meV.


We thus find a very good quantitative agreement with the aforementioned experimental values for the gaps of the $n=2$ and $n=3$ modes without further fitting parameters.  The success of this purely 1D theory is also consistent with the observation that these higher order modes are dispersionless along both the interchain and interplane directions. 

A physical interpretation of the bound states is as follows.  The
elementary excitations of the system (which comprise the $E_1$ mode)
are magnons, which behave as massive relativistic particles.  These
come in two ``flavors'', with spin $S^z=\pm 1$ along the Ising axis.
The $E_2$ mode arises as a bound state of two unlike particles, and
hence carries $S^z=0$, which is why it is longitudinal.  The third
$E_3$ mode corresponds to a bound state of two like and one unlike
particles, e.g. two $S^z=+1$ and one $S^z=-1$ magnons (or vice-versa),
which gives it a total $S^z=\pm 1$ implying it is transverse.  The
analytic treatment in the strict $1/S$ expansion discussed in the SM
confirms the presence of these bound states, though the DMRG
calculations are more quantitative.  Multiple magnon binding in
NaMnO$_2$ is therefore a remarkable and unexpected manifestation of
new physics accessible in higher spin one dimensional chains.


In summary, we have shown that the quasi-1D spin dynamics endemic to the anisotropic triangular lattice of NaMnO$_2$ manifest $n=2$ and $n=3$ high order magnon bound states. This result is captured by a semiclassical theory of interacting magnons within a 1D chain weakly bound by uniaxial single-ion anisotropy, which effectively maps to few body droplet models of interacting bosons bound via weak delta function potentials.  The result is a striking manifestation of strongly 1D quantum effects within the dynamics of a classical ($S=2$), planar antiferromagnet.  Therefore NaMnO$_2$, and we propose likely other anisotropic triangular antiferromagnets with weak, Ising-like single ion anisotropy, are appealing platforms for exploring few body interactions in a condensed matter setting.

\begin{acknowledgments}
This work was supported by DOE, Office of Science, Basic Energy Sciences under Award DE-SC0017752 (S.D.W., R.D. and M.B.). Work by L.B. was supported by the DOE, Office of Science, Basic Energy Sciences under Award No. DE-FG02-08ER46524. This research is funded in part by the Gordon and Betty Moore Foundation through Grant GBMF8690 to UCSB to support the work of A.K. A.J.R.H. thanks the Nanyang Technological University for financial support through the CN Yang Scholars Programme. M.B. also received partial support from the National Science Foundation Graduate Research Fellowship Program under Grant No. 1650114.  Use was made of the computational facilities administered by the Center for Scientific Computing at the CNSI and MRL (an NSF MRSEC; DMR-1720256) and purchased through NSF CNS-1725797.  
\end{acknowledgments}

\bibliography{bibliography}

\begin{thebibliography}{25}
\expandafter\ifx\csname natexlab\endcsname\relax\def\natexlab#1{#1}\fi
\expandafter\ifx\csname bibnamefont\endcsname\relax
  \def\bibnamefont#1{#1}\fi
\expandafter\ifx\csname bibfnamefont\endcsname\relax
  \def\bibfnamefont#1{#1}\fi
\expandafter\ifx\csname citenamefont\endcsname\relax
  \def\citenamefont#1{#1}\fi
\expandafter\ifx\csname url\endcsname\relax
  \def\url#1{\texttt{#1}}\fi
\expandafter\ifx\csname urlprefix\endcsname\relax\def\urlprefix{URL }\fi
\providecommand{\bibinfo}[2]{#2}
\providecommand{\eprint}[2][]{\url{#2}}

\bibitem[{\citenamefont{Haldane and Zirnbauer}(1993)}]{haldane1993exact}
\bibinfo{author}{\bibfnamefont{F.}~\bibnamefont{Haldane}} \bibnamefont{and}
  \bibinfo{author}{\bibfnamefont{M.}~\bibnamefont{Zirnbauer}},
  \bibinfo{journal}{Physical review letters} \textbf{\bibinfo{volume}{71}},
  \bibinfo{pages}{4055} (\bibinfo{year}{1993}).

\bibitem[{\citenamefont{Eggert et~al.}(1994)\citenamefont{Eggert, Affleck, and
  Takahashi}}]{PhysRevLett.73.332}
\bibinfo{author}{\bibfnamefont{S.}~\bibnamefont{Eggert}},
  \bibinfo{author}{\bibfnamefont{I.}~\bibnamefont{Affleck}}, \bibnamefont{and}
  \bibinfo{author}{\bibfnamefont{M.}~\bibnamefont{Takahashi}},
  \bibinfo{journal}{Phys. Rev. Lett.} \textbf{\bibinfo{volume}{73}},
  \bibinfo{pages}{332} (\bibinfo{year}{1994}),
  \urlprefix\url{https://link.aps.org/doi/10.1103/PhysRevLett.73.332}.

\bibitem[{\citenamefont{Haldane}(1983{\natexlab{a}})}]{PhysRevLett.50.1153}
\bibinfo{author}{\bibfnamefont{F.~D.~M.} \bibnamefont{Haldane}},
  \bibinfo{journal}{Phys. Rev. Lett.} \textbf{\bibinfo{volume}{50}},
  \bibinfo{pages}{1153} (\bibinfo{year}{1983}{\natexlab{a}}),
  \urlprefix\url{https://link.aps.org/doi/10.1103/PhysRevLett.50.1153}.

\bibitem[{\citenamefont{Affleck}(1989)}]{affleck1989quantum}
\bibinfo{author}{\bibfnamefont{I.}~\bibnamefont{Affleck}},
  \bibinfo{journal}{Journal of Physics: Condensed Matter}
  \textbf{\bibinfo{volume}{1}}, \bibinfo{pages}{3047} (\bibinfo{year}{1989}).

\bibitem[{\citenamefont{Haldane}(1983{\natexlab{b}})}]{haldane1983continuum}
\bibinfo{author}{\bibfnamefont{F.~D.~M.} \bibnamefont{Haldane}},
  \bibinfo{journal}{Physics Letters A} \textbf{\bibinfo{volume}{93}},
  \bibinfo{pages}{464} (\bibinfo{year}{1983}{\natexlab{b}}).

\bibitem[{\citenamefont{Mattis}(1986)}]{RevModPhys.58.361}
\bibinfo{author}{\bibfnamefont{D.~C.} \bibnamefont{Mattis}},
  \bibinfo{journal}{Rev. Mod. Phys.} \textbf{\bibinfo{volume}{58}},
  \bibinfo{pages}{361} (\bibinfo{year}{1986}),
  \urlprefix\url{https://link.aps.org/doi/10.1103/RevModPhys.58.361}.

\bibitem[{\citenamefont{Sekino and Nishida}(2018)}]{PhysRevA.97.011602}
\bibinfo{author}{\bibfnamefont{Y.}~\bibnamefont{Sekino}} \bibnamefont{and}
  \bibinfo{author}{\bibfnamefont{Y.}~\bibnamefont{Nishida}},
  \bibinfo{journal}{Phys. Rev. A} \textbf{\bibinfo{volume}{97}},
  \bibinfo{pages}{011602} (\bibinfo{year}{2018}),
  \urlprefix\url{https://link.aps.org/doi/10.1103/PhysRevA.97.011602}.

\bibitem[{\citenamefont{Zaccanti et~al.}(2009)\citenamefont{Zaccanti, Deissler,
  D’Errico, Fattori, Jona-Lasinio, M{\"u}ller, Roati, Inguscio, and
  Modugno}}]{zaccanti2009observation}
\bibinfo{author}{\bibfnamefont{M.}~\bibnamefont{Zaccanti}},
  \bibinfo{author}{\bibfnamefont{B.}~\bibnamefont{Deissler}},
  \bibinfo{author}{\bibfnamefont{C.}~\bibnamefont{D’Errico}},
  \bibinfo{author}{\bibfnamefont{M.}~\bibnamefont{Fattori}},
  \bibinfo{author}{\bibfnamefont{M.}~\bibnamefont{Jona-Lasinio}},
  \bibinfo{author}{\bibfnamefont{S.}~\bibnamefont{M{\"u}ller}},
  \bibinfo{author}{\bibfnamefont{G.}~\bibnamefont{Roati}},
  \bibinfo{author}{\bibfnamefont{M.}~\bibnamefont{Inguscio}}, \bibnamefont{and}
  \bibinfo{author}{\bibfnamefont{G.}~\bibnamefont{Modugno}},
  \bibinfo{journal}{Nature Physics} \textbf{\bibinfo{volume}{5}},
  \bibinfo{pages}{586} (\bibinfo{year}{2009}).

\bibitem[{\citenamefont{Sudan et~al.}(2009)\citenamefont{Sudan, L\"uscher, and
  L\"auchli}}]{PhysRevB.80.140402}
\bibinfo{author}{\bibfnamefont{J.}~\bibnamefont{Sudan}},
  \bibinfo{author}{\bibfnamefont{A.}~\bibnamefont{L\"uscher}},
  \bibnamefont{and} \bibinfo{author}{\bibfnamefont{A.~M.}
  \bibnamefont{L\"auchli}}, \bibinfo{journal}{Phys. Rev. B}
  \textbf{\bibinfo{volume}{80}}, \bibinfo{pages}{140402}
  (\bibinfo{year}{2009}),
  \urlprefix\url{https://link.aps.org/doi/10.1103/PhysRevB.80.140402}.

\bibitem[{\citenamefont{Kecke et~al.}(2007)\citenamefont{Kecke, Momoi, and
  Furusaki}}]{PhysRevB.76.060407}
\bibinfo{author}{\bibfnamefont{L.}~\bibnamefont{Kecke}},
  \bibinfo{author}{\bibfnamefont{T.}~\bibnamefont{Momoi}}, \bibnamefont{and}
  \bibinfo{author}{\bibfnamefont{A.}~\bibnamefont{Furusaki}},
  \bibinfo{journal}{Phys. Rev. B} \textbf{\bibinfo{volume}{76}},
  \bibinfo{pages}{060407} (\bibinfo{year}{2007}),
  \urlprefix\url{https://link.aps.org/doi/10.1103/PhysRevB.76.060407}.

\bibitem[{\citenamefont{Dally et~al.}(2018{\natexlab{a}})\citenamefont{Dally,
  Zhao, Xu, Chisnell, Stone, Lynn, Balents, and Wilson}}]{Dally_2018}
\bibinfo{author}{\bibfnamefont{R.~L.} \bibnamefont{Dally}},
  \bibinfo{author}{\bibfnamefont{Y.}~\bibnamefont{Zhao}},
  \bibinfo{author}{\bibfnamefont{Z.}~\bibnamefont{Xu}},
  \bibinfo{author}{\bibfnamefont{R.}~\bibnamefont{Chisnell}},
  \bibinfo{author}{\bibfnamefont{M.~B.} \bibnamefont{Stone}},
  \bibinfo{author}{\bibfnamefont{J.~W.} \bibnamefont{Lynn}},
  \bibinfo{author}{\bibfnamefont{L.}~\bibnamefont{Balents}}, \bibnamefont{and}
  \bibinfo{author}{\bibfnamefont{S.~D.} \bibnamefont{Wilson}},
  \bibinfo{journal}{Nat. Commun.} \textbf{\bibinfo{volume}{9}},
  \bibinfo{pages}{2188} (\bibinfo{year}{2018}{\natexlab{a}}),
  \urlprefix\url{https://doi.org/10.1038/s41467-018-04601-1}.

\bibitem[{\citenamefont{Giot et~al.}(2007)\citenamefont{Giot, Chapon,
  Androulakis, Green, Radaelli, and Lappas}}]{Giot_PRL_2007}
\bibinfo{author}{\bibfnamefont{M.}~\bibnamefont{Giot}},
  \bibinfo{author}{\bibfnamefont{L.~C.} \bibnamefont{Chapon}},
  \bibinfo{author}{\bibfnamefont{J.}~\bibnamefont{Androulakis}},
  \bibinfo{author}{\bibfnamefont{M.~A.} \bibnamefont{Green}},
  \bibinfo{author}{\bibfnamefont{P.~G.} \bibnamefont{Radaelli}},
  \bibnamefont{and} \bibinfo{author}{\bibfnamefont{A.}~\bibnamefont{Lappas}},
  \bibinfo{journal}{Phys. Rev. Lett.} \textbf{\bibinfo{volume}{99}},
  \bibinfo{pages}{247211} (\bibinfo{year}{2007}),
  \urlprefix\url{https://link.aps.org/doi/10.1103/PhysRevLett.99.247211}.

\bibitem[{\citenamefont{Zorko et~al.}(2008)\citenamefont{Zorko, El~Shawish,
  Ar\ifmmode~\check{c}\else \v{c}\fi{}on, Jagli\ifmmode \check{c}\else
  \v{c}\fi{}i\ifmmode~\acute{c}\else \'{c}\fi{}, Lappas, van Tol, and
  Brunel}}]{PhysRevB.77.024412}
\bibinfo{author}{\bibfnamefont{A.}~\bibnamefont{Zorko}},
  \bibinfo{author}{\bibfnamefont{S.}~\bibnamefont{El~Shawish}},
  \bibinfo{author}{\bibfnamefont{D.}~\bibnamefont{Ar\ifmmode~\check{c}\else
  \v{c}\fi{}on}}, \bibinfo{author}{\bibfnamefont{Z.}~\bibnamefont{Jagli\ifmmode
  \check{c}\else \v{c}\fi{}i\ifmmode~\acute{c}\else \'{c}\fi{}}},
  \bibinfo{author}{\bibfnamefont{A.}~\bibnamefont{Lappas}},
  \bibinfo{author}{\bibfnamefont{H.}~\bibnamefont{van Tol}}, \bibnamefont{and}
  \bibinfo{author}{\bibfnamefont{L.~C.} \bibnamefont{Brunel}},
  \bibinfo{journal}{Phys. Rev. B} \textbf{\bibinfo{volume}{77}},
  \bibinfo{pages}{024412} (\bibinfo{year}{2008}),
  \urlprefix\url{https://link.aps.org/doi/10.1103/PhysRevB.77.024412}.

\bibitem[{\citenamefont{Stock et~al.}(2009)\citenamefont{Stock, Chapon,
  Adamopoulos, Lappas, Giot, Taylor, Green, Brown, and
  Radaelli}}]{stock2009one}
\bibinfo{author}{\bibfnamefont{C.}~\bibnamefont{Stock}},
  \bibinfo{author}{\bibfnamefont{L.}~\bibnamefont{Chapon}},
  \bibinfo{author}{\bibfnamefont{O.}~\bibnamefont{Adamopoulos}},
  \bibinfo{author}{\bibfnamefont{A.}~\bibnamefont{Lappas}},
  \bibinfo{author}{\bibfnamefont{M.}~\bibnamefont{Giot}},
  \bibinfo{author}{\bibfnamefont{J.}~\bibnamefont{Taylor}},
  \bibinfo{author}{\bibfnamefont{M.}~\bibnamefont{Green}},
  \bibinfo{author}{\bibfnamefont{C.}~\bibnamefont{Brown}}, \bibnamefont{and}
  \bibinfo{author}{\bibfnamefont{P.}~\bibnamefont{Radaelli}},
  \bibinfo{journal}{Physical review letters} \textbf{\bibinfo{volume}{103}},
  \bibinfo{pages}{077202} (\bibinfo{year}{2009}).

\bibitem[{\citenamefont{Dally et~al.}(2017)\citenamefont{Dally, Clément,
  Chisnell, Taylor, Butala, Doan-Nguyen, Balasubramanian, Lynn, Grey, and
  Wilson}}]{DALLY2017203}
\bibinfo{author}{\bibfnamefont{R.}~\bibnamefont{Dally}},
  \bibinfo{author}{\bibfnamefont{R.~J.} \bibnamefont{Clément}},
  \bibinfo{author}{\bibfnamefont{R.}~\bibnamefont{Chisnell}},
  \bibinfo{author}{\bibfnamefont{S.}~\bibnamefont{Taylor}},
  \bibinfo{author}{\bibfnamefont{M.}~\bibnamefont{Butala}},
  \bibinfo{author}{\bibfnamefont{V.}~\bibnamefont{Doan-Nguyen}},
  \bibinfo{author}{\bibfnamefont{M.}~\bibnamefont{Balasubramanian}},
  \bibinfo{author}{\bibfnamefont{J.~W.} \bibnamefont{Lynn}},
  \bibinfo{author}{\bibfnamefont{C.~P.} \bibnamefont{Grey}}, \bibnamefont{and}
  \bibinfo{author}{\bibfnamefont{S.~D.} \bibnamefont{Wilson}},
  \bibinfo{journal}{Journal of Crystal Growth} \textbf{\bibinfo{volume}{459}},
  \bibinfo{pages}{203 } (\bibinfo{year}{2017}), ISSN \bibinfo{issn}{0022-0248},
  \urlprefix\url{http://www.sciencedirect.com/science/article/pii/S0022024816308582}.

\bibitem[{\citenamefont{Granroth et~al.}(2010)\citenamefont{Granroth,
  Kolesnikov, Sherline, Clancy, Ross, Ruff, Gaulin, and
  Nagler}}]{Granroth_2010}
\bibinfo{author}{\bibfnamefont{G.~E.} \bibnamefont{Granroth}},
  \bibinfo{author}{\bibfnamefont{A.~I.} \bibnamefont{Kolesnikov}},
  \bibinfo{author}{\bibfnamefont{T.~E.} \bibnamefont{Sherline}},
  \bibinfo{author}{\bibfnamefont{J.~P.} \bibnamefont{Clancy}},
  \bibinfo{author}{\bibfnamefont{K.~A.} \bibnamefont{Ross}},
  \bibinfo{author}{\bibfnamefont{J.~P.~C.} \bibnamefont{Ruff}},
  \bibinfo{author}{\bibfnamefont{B.~D.} \bibnamefont{Gaulin}},
  \bibnamefont{and} \bibinfo{author}{\bibfnamefont{S.~E.}
  \bibnamefont{Nagler}}, \bibinfo{journal}{Journal of Physics: Conference
  Series} \textbf{\bibinfo{volume}{251}}, \bibinfo{pages}{012058}
  (\bibinfo{year}{2010}),
  \urlprefix\url{https://doi.org/10.1088%2F1742-6596%2F251%2F1%2F012058}.

\bibitem[{\citenamefont{Ewings et~al.}(2016)\citenamefont{Ewings, Buts, Le, van
  Duijn, Bustinduy, and Perring}}]{Horace}
\bibinfo{author}{\bibfnamefont{R.}~\bibnamefont{Ewings}},
  \bibinfo{author}{\bibfnamefont{A.}~\bibnamefont{Buts}},
  \bibinfo{author}{\bibfnamefont{M.}~\bibnamefont{Le}},
  \bibinfo{author}{\bibfnamefont{J.}~\bibnamefont{van Duijn}},
  \bibinfo{author}{\bibfnamefont{I.}~\bibnamefont{Bustinduy}},
  \bibnamefont{and} \bibinfo{author}{\bibfnamefont{T.}~\bibnamefont{Perring}},
  \bibinfo{journal}{Nuclear Instruments and Methods in Physics Research Section
  A: Accelerators, Spectrometers, Detectors and Associated Equipment}
  \textbf{\bibinfo{volume}{834}}, \bibinfo{pages}{132 } (\bibinfo{year}{2016}),
  ISSN \bibinfo{issn}{0168-9002},
  \urlprefix\url{http://www.sciencedirect.com/science/article/pii/S016890021630777X}.

\bibitem[{\citenamefont{Dally et~al.}(2018{\natexlab{b}})\citenamefont{Dally,
  Chisnell, Harriger, Liu, Lynn, and Wilson}}]{Dally_2018_PRB}
\bibinfo{author}{\bibfnamefont{R.~L.} \bibnamefont{Dally}},
  \bibinfo{author}{\bibfnamefont{R.}~\bibnamefont{Chisnell}},
  \bibinfo{author}{\bibfnamefont{L.}~\bibnamefont{Harriger}},
  \bibinfo{author}{\bibfnamefont{Y.}~\bibnamefont{Liu}},
  \bibinfo{author}{\bibfnamefont{J.~W.} \bibnamefont{Lynn}}, \bibnamefont{and}
  \bibinfo{author}{\bibfnamefont{S.~D.} \bibnamefont{Wilson}},
  \bibinfo{journal}{Phys. Rev. B} \textbf{\bibinfo{volume}{98}},
  \bibinfo{pages}{144444} (\bibinfo{year}{2018}{\natexlab{b}}),
  \urlprefix\url{https://link.aps.org/doi/10.1103/PhysRevB.98.144444}.

\bibitem[{sup()}]{supplemental}
\bibinfo{note}{See supplemental information, which includes references
  \cite{White2008}}.

\bibitem[{\citenamefont{Schollw\"ock}(2011)}]{Schollwoeck2011}
\bibinfo{author}{\bibfnamefont{U.}~\bibnamefont{Schollw\"ock}},
  \bibinfo{journal}{Annals of Physics} \textbf{\bibinfo{volume}{326}},
  \bibinfo{pages}{96} (\bibinfo{year}{2011}).

\bibitem[{ITe()}]{ITensor}
\bibinfo{note}{ITensor Library,
  \href{http://itensor.org/}{http://itensor.org/}}.

\bibitem[{\citenamefont{White}(1992)}]{White1992}
\bibinfo{author}{\bibfnamefont{S.~R.} \bibnamefont{White}},
  \bibinfo{journal}{Phys. Rev. Lett.} \textbf{\bibinfo{volume}{69}},
  \bibinfo{pages}{2863} (\bibinfo{year}{1992}).

\bibitem[{\citenamefont{Vidal}(2003)}]{Vidal2003}
\bibinfo{author}{\bibfnamefont{G.}~\bibnamefont{Vidal}},
  \bibinfo{journal}{Phys. Rev. Lett.} \textbf{\bibinfo{volume}{91}},
  \bibinfo{pages}{147902} (\bibinfo{year}{2003}).

\bibitem[{\citenamefont{Vidal}(2004)}]{Vidal2004}
\bibinfo{author}{\bibfnamefont{G.}~\bibnamefont{Vidal}},
  \bibinfo{journal}{Phys. Rev. Lett.} \textbf{\bibinfo{volume}{93}},
  \bibinfo{pages}{040502} (\bibinfo{year}{2004}),
  \urlprefix\url{https://link.aps.org/doi/10.1103/PhysRevLett.93.040502}.

\bibitem[{\citenamefont{White and Affleck}(2008)}]{White2008}
\bibinfo{author}{\bibfnamefont{S.~R.} \bibnamefont{White}} \bibnamefont{and}
  \bibinfo{author}{\bibfnamefont{I.}~\bibnamefont{Affleck}},
  \bibinfo{journal}{Phys. Rev. B} \textbf{\bibinfo{volume}{77}},
  \bibinfo{pages}{134437} (\bibinfo{year}{2008}),
  \urlprefix\url{https://link.aps.org/doi/10.1103/PhysRevB.77.134437}.

\end{thebibliography}

\end{document}